# Learning about the Energy of a Hurricane System through an Estimation Epistemic Game

Bahar Modir*, Paul W. Irving†**, Steven F. Wolf**, and Eleanor C. Sayre*

*Department of Physics, Kansas State University, 116 Cardwell Hall, Manhattan, KS 66506
†Department of Physics and Astronomy, Michigan State University, 4261 Biomedical and Physical Sciences Building, East Lansing, MI 48824
**CREATE for STEM Institute, Michigan State University, 115 Erickson Hall, East Lansing, MI 48824

**Abstract:** As part of a study into students' problem solving behaviors, we asked upper-division physics students to solve estimation problems in clinical interviews. We use the Resources Framework and epistemic games to describe students' problem solving moves. We present a new epistemic game, the "estimation epistemic game". In the estimation epistemic game, students break the larger problem into a series of smaller, tractable problems. Within each sub-problem, they try to remember a method for solving the problem, and use estimation and reasoning abilities to justify their answers. We demonstrate how a single case study student plays the game to estimate the total energy in a hurricane. Finally, we discuss the implications of epistemic game analysis for other estimation problems.

**Keywords:** Estimation epistemic game, resources, hurricane problem.

**PACS:** 01.40.Fk

## INTRODUCTION

Solving problems plays a major role in studying physics. Various researchers have developed theories and strategies to study students' problem solving in different contexts. Estimation problems are a type of Fermi questions[1] whose exact solutions are too difficult to measure and they may not even be entirely precise. Solving Fermi questions requires intuition, mathematics, common sense, reasoning, and the skill to break down complex problems into discrete solvable parts. Estimation problems do not have a single exact route towards solution and that makes them ideal for studying students' problem solving decisions.

We are interested in understanding the mechanistic underpinnings of the problem solving approach through the lens of epistemic games (e-games): an activation of patterns of activities that can be associated with sets of cognitive resources.[2] Normative e-games were first proposed by Collins and Ferguson as a set of rules and strategies to analyze phenomena in terms of their structure, functions, and processes.[3] Tuminaro and Redish[2] identified six different descriptive e-games specific to physics rather than common to all scientific disciplines, and Chen[4] identified an answer-making e-game in students' solutions to conceptual problems.

We use the Resources Framework[5] to present an analysis of a single student's thinking at the micro-level. We are particularly interested in her problem solving moves and how she activates p-prims,[6] mathematical,[2] and procedural resources.[7] Students' epistemic stances are manifold and context-sensitive.[8] By applying e-games to estimation-type problems, we can describe students' ways of thinking and their knowledge and developmental strategies for finding a solution in context-dependent and manifold ways.

Through e-games and the Resources Framework, we can describe students' tacit expectations[2] about how to approach solving physics problems. The structure of an e-game consists of two components: the entry conditions and the moves. The entry conditions are determined by an individual's expectations about the particular situation or physics problem solving. The collection of qualitative and quantitative resources and reasoning abilities that an individual draws on while playing a particular e-game constitutes the knowledge base component of an e-game. The actual path that students follow during problem solving in physics varies from problem-to-problem and student-to-student.[9] A single student will take different paths depending on the relative difficulty of the problem and which resources they activate.[4]

In this paper, we identify a new e-game: the "estimation epistemic game" (e-e-game), which involves new moves, entry, and exit conditions. In it, the student activates sets of resources and applies an estimation approach to evaluate possible solutions and produce knowledge and arguments.

## METHODOLOGY

We interviewed "Ava", an upper-division physics major, using a talk-aloud protocol. She solved the following problem: "Estimate the total energy in a typical hurricane system. State explicitly the assumptions that you make, explain your reasoning, and assess your result." Ava had access to a paper sheet with unit conversions and useful constants. Ava plays the game in about 14 minutes (in three more minutes, she answers an unrelated question asked by the interviewer).

Our focus for identifying the new e-e-game played by Ava is based on the observation of specific types of declarative and procedural resources activated in association with specific moves. We used micro-genetic analysis as a comprehensive way to examine moment-by-moment conceptual changes of a student's learning activity. This method analyzes students' discourse and physical activities within a short period of time.[10] In order to observe moves in the macro level of the game, we need to identify and enlarge fine grained information underlying sub-moves through the lens of a micro-analysis method. The ability to understand macro-level changes of developmental time is related to observing and discovering micro-level changes of real time.[10] The smallest observable time scales we have seen in the activation of resources were in the order of one second.

Based on the micro-genetic analysis of the student's behavior, our observations, and evidence from previous studies, we propose that Ava is entering a previously un-articulated e-game, which consists of several basic moves.

## GAME STRUCTURE

The goal of the e-e-game is to solve a problem through estimation. There are six basic moves in this game (see Fig. 1.): 1. problematize, 2. propose method, 3. what to remember, 4. see if parts are enough, 5. pure calculations and 6. evaluations. Ava enters the game by being confronted with the estimation problem. If Ava takes a particular path through the activated resources and refers to that specific path several times, we can call this path a part of an e-e-game or a move. During the first move, Ava starts with a quick, intuitive response. In the second move, a consistent method based on the expectation of Ava in the initial move is proposed. In the third move, a method is related to the physical concepts by written equations. The goal of the fourth move is to remember different resources in order to estimate a numerical answer for all of the physical variables. In the fifth move, by using mathematical calculations, Ava finds an answer and exits the game after evaluating her answer.

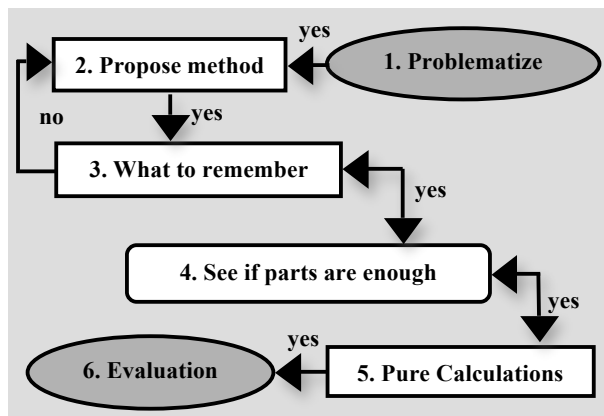

**FIGURE 1.** Schematic graph of the e-e-game played by the student.

### 1. Problematize

In this case study, Ava enters the game by reading the first sentence of the question. She finds the question hard and begins to activate resources based on her prior informal knowledge about hurricanes, to obtain a general idea about the nature of the problem.

*Ava: (00:08) Oh this sounds horrid! So what is a hurricane? A hurricane is mostly air that's moving in some sort of rotational system. Suppose… um… I suppose it's gonna (sic) lot to do with more high and low pressure areas.*

Initially, she problematizes the hurricane system by describing a "rotational system" and "high and low pressure areas."

### 2. Propose Method

After she problematizes the question, Ava begins the second move, which is selecting a method. In this section, Ava starts with choosing the rotational energy formula, which is resulted from her expectations of the physics concepts in the question.

*Ava: (00:52) Ok. I'm gonna start with just finding the rotational energy of the hurricane system. Because I think that sounds like something I can do <laughing> pretty reasonably. Ah, ok …so rotational energy.*

To calculate the inertia of a hurricane system, Ava begins to activate the physics quantity of inertia of a point mass and links it to other mathematical resources.

## 3. What To Remember

In this part Ava tabulates the facts, concepts, and equations by activating declarative resources and asking herself whether she can move forward. However, she doesn't remember enough about rotational systems to move forward with this method. She returns to the prior move, proposing a new method: examining the kinetic energy of the system.

*Ava: (04:50) Energy equals $1/2\ mv^2$. So the mass is going to be the density times the volume.*

While she records the related equation consistent with her method, new resources come to Ava's mind. This allows her to define new sub-targets and use correlative strategies to assemble and assess if the parts are sufficient to confer a reasonable estimation. Then she proceeds through the game and connects different parts of the problem.

## 4. See If Parts Are Enough

Ava activates her procedural resources by asking questions about how to solve different aspects of the problem: a process of mechanistic reasoning. In this step her resources become more developed by supporting them through reasonable estimations, examinations, measurements, and scales. Ava utilizes her conversion sheet to quickly determine that the density of water is greater than air. Within this short 'micro moment' she attempts to find a reason to neglect the effect of the water density by considering that the droplets of water consume less space than air.

*Ava: (05:12) We will say it's probably mostly air. Water is significantly higher density but water droplets are small compared to the total volume of air. Maybe I should lowball one of these estimations and go on the other side of my other estimates. <pause> Ok ... how big is our hurricane? They usually look like this <showing with hands> around Florida - about that big. Let's see just a (sic) order of magnitude ... How big should this radius be? I'm assuming it's a cylinder because that sounds reasonable.*

Ava uses her balancing resources (*"lowball one…go on the other side"*) to argue that she doesn't need to take into account the density of water. Then she activates her size resources to help her make an estimation to relate the size of a typical hurricane to the size of Florida. She decides that the shape of a typical hurricane is a cylinder (*"I assume…"*). We infer that she chooses this shape (*"that sounds reasonable"*) because she wants to perform some possible quick calculations (*"how big should this radius be?"*). We infer she has linked the intuition to her geometrical resources and specified the volume as a cylinder.

## 5. Pure Calculations

Now it's time to conduct her math calculations. At this point, by getting close to the end of the game, she increases her speed. Her affect becomes markedly more positive, and she enjoys this part of the game.

## 6. Evaluations

Before exiting the game, Ava does a kind of evaluation of her number by checking the units and considering her solution has some errors and might be different from the actual energy in a typical hurricane. In contrast to her earlier work, which is rife with sense-making and mechanistic reasoning, here Ava bemoans her inability to make sense of her number.

*Ava: (15:40) We will say, ten to the fifteenth joules for a typical hurricane system. I don't even know if that makes sense. I don't have a good way to check that.*

## Patterns of Moves

Ava is unable to solve the estimation problem at a glance; she breaks down the main question into several smaller estimation problems to generate heuristics sufficient for her to navigate sub-problems and arrive at a temporary goal. Ava is now in a position to use the fifth move (pure calculation) to combine the sub-targets and find the main answer. She switches several times between her third and fourth moves, mapping physical concepts and plugging estimate-based numbers into equations. She stops alternating when she has figured out all of the unknown physical quantities (Fig. 2). In the initial eight minutes of problem solving, Ava has longer, deep thinking periods while mostly inter-playing between the third and fourth moves. In the second time duration of six minutes of the game, she is using the third, fourth, and fifth moves, during which the rate of the changes in moves has increased thus speeding up the game.

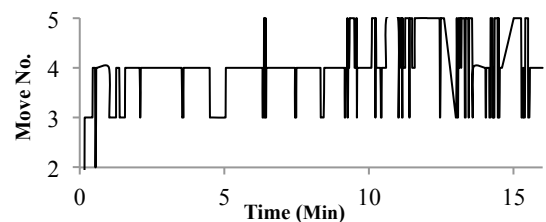

**FIGURE 2.** Ava plays the e-e-game. Notice that initially she spends more time in each move, but around minute 10, she starts to change rapidly between three moves.

In Table 1, we have indicated the total time that has been spent in each of these moves. Most of her time has been spent in the fourth move within which she does most of her estimations. She refers back to move 3 the largest number of times (34 times), each time for a brief period before returning to move 4 (in the first 8 minutes) or moves 4 and 5 (in the last 7 minutes).

TABLE 1. Times and recall numbers allocated to different moves of the e-e-game.

| Move number | 3 | 4 | 5 |
|---|---|---|---|
| Minutes in move | 1.86 | 5.41 | 2.29 |
| # transitions to here | 34 | 27 | 26 |

## DISCUSSION AND IMPLICATIONS

This game has some moves that are both similar to and different from the e-games identified by Tuminaro and Redish[2] and the AMEG identified by Chen et al.[4] The fourth move is similar to AMEG as students justify their answers by providing reasonable assumptions. This is similar to the justification path of AMEG, in which students try to build a justification for their remembered or intuitive answer. The third move of this game is comparable to the second part of the Mapping Mathematics to Meaning e-game. In both, students are relating the target to other concepts by plugging them into an equation, but it is different from the third part of the Mapping Mathematics to Meaning e-game which is telling a story. Another likeness is with the sub-part of the third step in the Recursive Plug-and-Chug e-game in identifying other unknown sub-target quantities rather than the target quantity; however, the difference is that students achieve a numerical answer without making sense out of that answer which differs from the reasoning conducted in the fourth move of the current study.

The e-e-game played by Ava with specific paths and moves might be an artifact derived from the estimation-based question. Given our data, we can't distinguish between Ava's specific game-playing details in this context and a more generalized characterization of the e-e-game. Her experience as an upper-level undergraduate student would likely affect her choice to enter a flexible e-e-game, with several sub-moves to activate a variety of specific relevant resources available to her. There may be a large overlap in students' activation of resources within both physical and mathematical domains. However, the order of activation in a specific game could lead to different paths resulting in numerous starting conditions, moves, and end outcomes. For future studies, we are interested in investigating in a larger scope other outcomes of the e-e-games that attempt to solve the same class of estimation problems.

## CONCLUSION

An e-game can be described as a list of activated procedural linked resources. By investigating the behavior of Ava at a finely grained level as she attempts to solve an estimation-based problem, we found that she enters into a previously unarticulated e-game by defining several sub-estimation problems. She activates different types of resources and maps them directly to each sub-problem situation, then combines the individual pieces of the sub-problems to find the final answer. Ava supports and links different resources mainly by using estimations in the form of an e-e-game. Physicists often use estimations to make a large number of difficult-to-solve problems tractable. It is worthwhile for students to solve estimation physics problems to develop practical, critical, and logical skills. Via e-e-games, students are able to apply their intuition and accessible knowledge in taking the first crucial steps of solving problems as a physicist.

## ACKNOWLEDGMENTS

We thank Dr. Sanjay Rebello, Dr. Brian O'Shea, Alanna Pawlak, and the K-State Writing Center. This work was supported by the K-State University Physics Department and Lyman Briggs College.